\documentclass[12pt,preprint]{aastex}
\usepackage{graphicx,epsfig,epsf,amssymb}      

\newcommand{\myemail}{litp@tsinghua.edu.cn}
\newcommand{\liuemail}{liuhao@ihep.ac.cn}
\def \<{\langle}
\def \>{\rangle}

\newcommand{\degree}{^\circ}

\begin{document}

  \title{Systematic Distortion in Cosmic Microwave Background Maps}
\author{Hao LIU\altaffilmark{1,3} \& Ti-Pei LI\altaffilmark{1,2}}
 \altaffiltext{1}{Key Lab. of Particle Astrophys., Inst. of High Energy Phys.,
Chinese Academy of Sciences, Beijing; \liuemail}
\altaffiltext{2}{Center for
Astrophysics, Tsinghua University, Beijing, China; \myemail}
\altaffiltext{3}{Graduate School of Chinese Academy of Sciences, Beijing}

\begin{abstract}
To minimize instrumentally induced systematic errors,
cosmic microwave background (CMB) anisotropy experiments measure
temperature differences across the sky using paires of horn antennas,
temperature map is recovered from temperature
differences obtained in sky survey through a map-making procedure.
To inspect and calibrate residual systematic errors in  recovered temperature maps
is important as
most previous studies of cosmology are based on these maps.
By analyzing pixel-ring couping and latitude dependence of CMB temperatures,
we find notable systematic deviation from  CMB Gaussianity
in released Wilkinson Microwave Anisotropy Probe (WMAP) maps.
The detected deviation can not be explained by the best-fit $\Lambda$CDM
cosmological model at a confidence level above $99\%$ 
and can not be ignored for a precision cosmology study.
\end{abstract}


\section{Introduction: differential observation and scan-ring}
To minimize instrumentally induced systematic errors,
the WMAP mission measures temperature differences across
the sky using paires of horn antennas with a fixed separation angle $\theta$,
temperature maps are recovered from raw time-ordered temperature
differences obtained in sky survey with a map-making algorithm$^{[1]}$.
When an antenna points to a sky pixel $i$, the scan path of the other one will
draw a ring $R_\theta(i)$ in the sky with angular radius  $\theta$
to the center pixel $i$.
The map temperature $t(i)$ of a pixel $i$
should be in some extent correlated to measured temperatures in its scan-ring
$R_\theta(i)$ through the map-making procedure.
The beam separation angle of WMAP radiometers  is $\theta\sim141\degree$.
The off-diagonal terms of noise covariance matrixes have been inspected
by the WMAP team  with two-point correlation functions and small positive blips
of order $0.3\%$ of the diagonal elements have been found
at $141\degree$ pixel separation angle$^{[2,3]}$.

Instead of the pixel-pixel noise coupling, in this work we inspect the
pixel-ring temperature coupling,
and find in released WMAP CMB maps that notable temperature distortions
exist on scan-rings of hot foreground sources: scan-rings of hot sources
are significantly cooled (presented in \S2) and strongest anti-correlations
between pixel and scan-ring temperatures appear at a separation angle
 $\theta\sim141\degree$ (\S3).
In \S4 we find a systematic deviation from CMB isotropy in
Galactic latitude distributions of released WMAP maps, which
can be understood by pixel-ring temperature coupling found in \S2 and \S3.
Finally we give a brief discussion in \S5.

\section{Cold scan-rings of hot sources}
\begin{figure}
\label{rings}
    \begin{center}
    \includegraphics[width=46mm,angle=90]{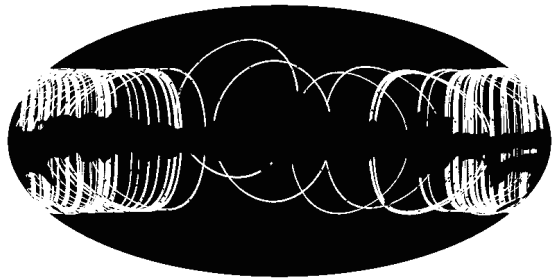}
\vspace{-3mm}   \caption{\footnotesize The scan-rings of 2000 hottest pixels in
the Q-band WMAP5 map in Galactic coordinates after using the Kp0 mask.
The ring width is $0.5\degree$.
}
    \end{center}
\vspace{-4mm}
\end{figure}

We choose 2000 brightest pixels that are close to the galactic plane
from five year WMAP (WMAP5) intensity maps$^{[4]}$ with HEALPix resolution parameter
$N_{side}=512$$^{[5]}$ and almost the same for different bands,
and draw their scan-rings of $0.5\degree$ width, shown
in Fig.~\ref{rings}. The Kp0 mask is used to exclude bright galactic sources.
All the scan-rings in Fig. \ref{rings} form a sample (shortened as
``sample 1'' henceforth) that covers about $15\%$
of the entire celestial sphere. Generally speaking, sample 1 is large enough
to merge most CMB anisotropy and the expectation of average temperature should be very close
to zero.

For the WMAP5 Q-, V- and W- band foreground-reduced maps and ILC map$^{[6]}$
we calculate the average temperature of sample 1 (``ring temperature'' hereafter)
respectively.
The results are listed in Table~1.
To estimate their significances we simulate observed CMB temperature maps
with the synfast program in HEALPix software package (available at
 http://healpix.jpl.nasa.gov) from the best fit $\Lambda$CDM model
power spectrum with proper beam function
and noise property. For each input map, 1000 simulated maps are created and their
ring temperatures calculated. The average ring temperature and its standard deviation
is shown in the bottom row of Table~1 for each input map.
From this table we can see that, for the four maps the ring temperatures
are all notably lower than the expectation with a confidence level higher than $99\%$.
With these results, it is obvious that map temperatures on scan-rings of hot sources
suffer systematical distortion, and the observed violence to CMB Gaussianity
should come, at least partially, from the combined effect of hot foreground sources
and WMAP's differential nature.

We have also rotated WMAP5 ILC temperature map (by rotating the north pole to other pixels) 
to place sample 1 to other locations and compute the new average temperatures. 
The original sample 1 is known to be abnormal; therefore it is excluded together 
with the KP0 foreground mask. The objective of this test is to check the
variances given in Table~1 with the original map, and the result should be close 
to simulation; otherwise the simulation is suspicious. The north pole of WMAP5 ILC 
map is rotated to all 3072 sky pixels in HEALPix
resolution $N_{side}=16$ and the expectation and RMS fluctuation 
of the average temperature of sample 1 are -0.5 $\mu$K and 4.3 $\mu$K respectively. 
This is consistent with Table~1; therefore, the simulation and
Table~1 are reliable.

\begin{table}
\begin{center}
\caption{Temperatures ($\mu$K) averaged over scan-rings of hot sources}
\vspace{1mm}
\begin{tabular}{ l l l l l }  \hline
   & Q-band & V-band & W-band & ILC map \\ \cline{2-5}
WMAP5   & -11.67   & -12.62 & -12.92 & -11.34 \\
 Simulation  & $0.09\pm 4.7$  & $-0.09\pm4.7$ & $0.09\pm4.8$ & $0.04\pm4.4$\\
Significance& $2.5\sigma$     &$2.7\sigma$& $2.7\sigma$&$2.7\sigma$\\
\hline
\end{tabular}
\end{center}
\end{table}

\section{Pixel-ring temperature coupling}
We use $t_R(i,\theta)$ to denote the average temperature over the scan ring
of a pixel $i$ with a separation angle $\theta$, i.e.
$t_R(i,\theta)=\<t(k)\>$ where $k\in R_\theta(i)$.
For a temperature map with
$N_{side}=128$, we calculate the cross-correlation coefficient
$C(\theta)$ between $t(i)$ and $t_R(i,\theta)$
for a designed separation angle $\theta$
\[ C(\theta)=\frac{\sum_i(t(i)-\overline{t})(t_R(i,\theta)-\overline{t}_R)}
{\sqrt{\sum_i(t(i)-\overline{t})^2(t_R(i,\theta)-\overline{t}_R)^2}}~.\]

\begin{figure}
\label{correlation}
   \begin{center}
\vspace{-1mm}\includegraphics[height=4.5cm, width=6.5cm, angle=0]{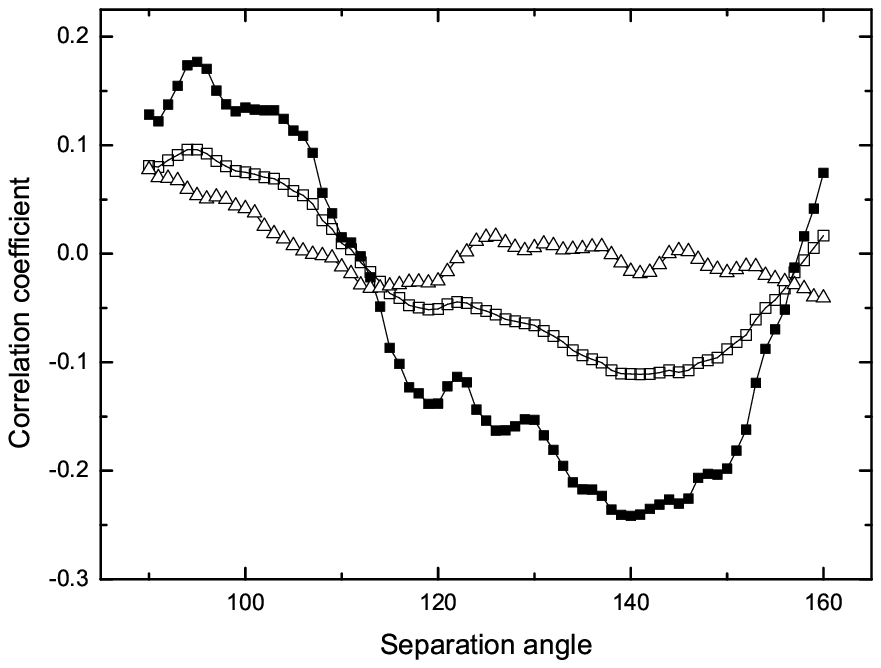}\\
\vspace{0mm}
\includegraphics[height=4.5cm, width=6.5cm, angle=0]{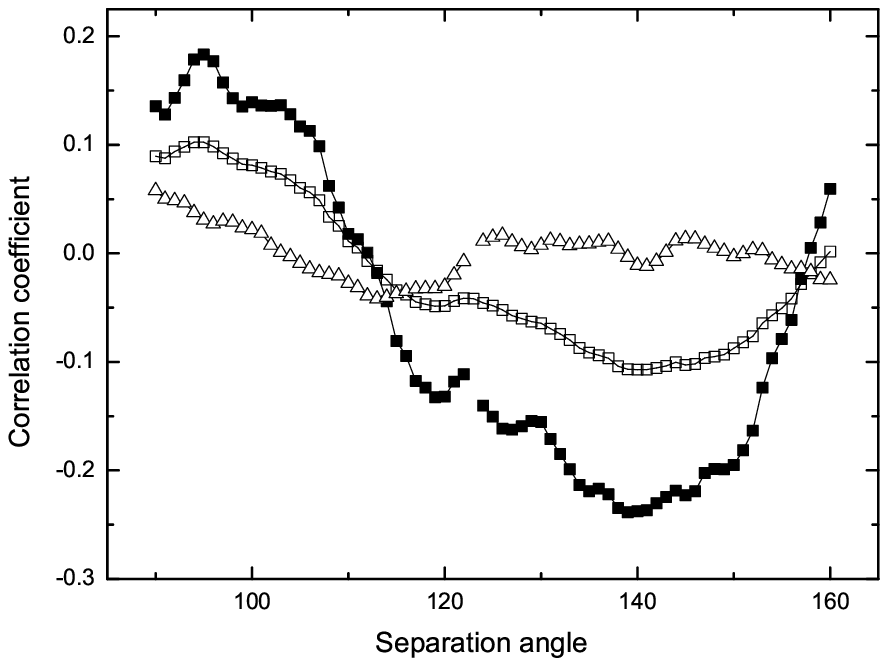}\\
\vspace{0mm}
 \hspace{1mm}\includegraphics[height=4.5cm, width=6.5cm, angle=0]{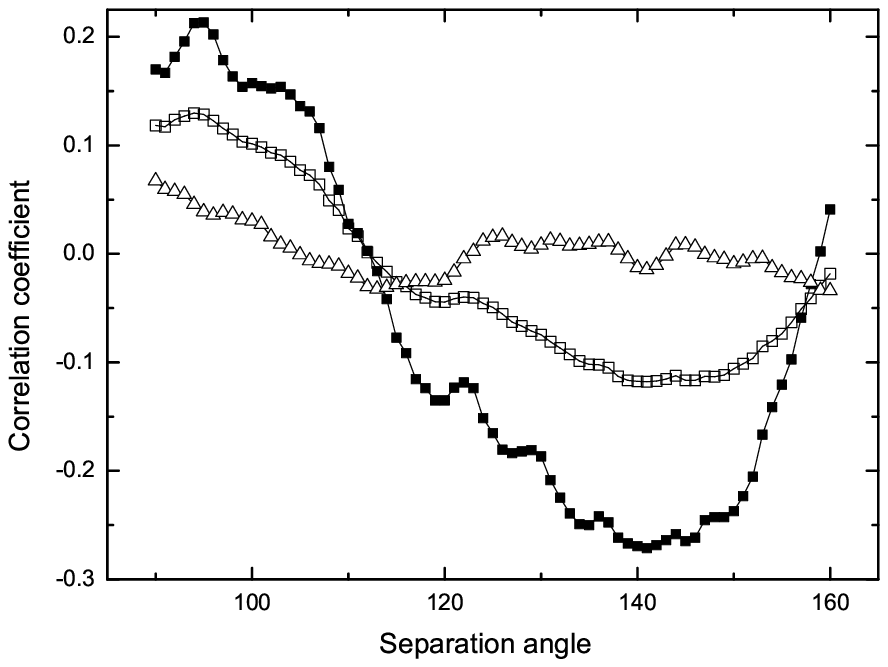}
\vspace{-3mm} \caption{\footnotesize Separation angle dependence of pixel-ring 
temperature correlation. The vertical coordinate shows the correlation coefficient 
between temperature $t(i)$ of pixel $i$ and average temperature $t_R(i,\theta)$ 
of the ring with separation angle $\theta$ to $i$. The abscissa marks 
the separation angle $\theta$ in degree.
{\it Filled square}: pixel $i$ only within the foreground mask Kp12.
{\it Square}: for whole sky. 
{\it Triangle}: for the sky region out of the mask Kp0.
{\sl Top panel}: WMAP5 Q-band.
{\sl Middle panel}: WMAP5 V-band.
{\sl Bottom panel}: WMAP5 W-band.  } 
   \end{center}
\vspace{-4mm}
\end{figure}

First, we limit pixel $i$ only within the region of foreground
mask Kp12 where contain hottest foreground sources and calculate
$C(\theta)$ at different $\theta$ and for
WMAP5 Q, V and W band maps separately. The obtained correlation distributions
are shown by filled squares in Fig.~2, where the strongest negative correlation
appear around $141\degree$ separation for each band.
The separation angle dependence of pixel-ring
temperature correlation is also obtained from the WMAP5 ILC map, which is
similar with what from the WMAP5 Q, V and W band maps shown in Fig.~2.
The correlation coefficients at $141\degree$ separation, $C(\theta=141\degree)$,
 from WMAP5 Q-, V- and W-band maps
are listed in Table~2.  The bottom row of Table~2 comes from 1000 simulations
for each band. The foreground emission maps$^{[6]}$
have to be used in the simulation calculation because we need the original recovered
temperatures (without foreground reduction) for $t(i)$, which is the reason that we do not
give the result for ILC map in Table~2 (ILC
map does not have corresponding foreground emission map).
From Table~2 we can see that the negative correlation
at $\theta=141\degree$ has a significance from $2.2\sigma$ to $3.9\sigma$.

\begin{table}
\begin{center}
\caption{Correlation coefficients
 between temperatures of pixels
within Kp12 mask and average temperatures on their $141\degree$ scan-rings}
\vspace{2mm}
\begin{tabular}{ l l l l }  \hline
   & Q-band & V-band & W-band  \\ \cline{2-4}
WMAP5   & -0.234   & -0.223 & -0.262 \\
 Simulation  & $0.001\pm 0.106$  & $-0.001\pm0.101$ & $-0.002\pm 0.066$ \\
Significance & $2.2\sigma$ & $2.2\sigma$ &$3.9\sigma$\\
\hline
\end{tabular}
\end{center}
\end{table}

In the sky region out of the Galaxy plane (out of Kp0 mask) the correlation dip
around the $141\degree$ separation almost completely disappear (shown by triangles
in Fig.~2), indicating again that the detected correlation structure
is most possibly a combined effect of WMAP differential observation and Galactic hot emission.
The broad feature of the correlation dip around the $141\degree$ separation
can be caused by finite width and noncircular response of instrument beam,
structure of hot emission regions, and diffusion of temperature distortion
in map-making process.
The anti-correlation between temperatures of Galactic plane and its
$141\degree$ scan-rings is consistent with the detected fact that
$141\degree$ scan-rings of hot sources being cold (shown in \S2).

Since the noise in WMAP temperature maps is known to be correlated 
at $141\degree$$^{[2,3]}$, it is necessary to check 
pixel-ring correlation with the same noise maps they used 
(e.g., Q1-Q2, V1-V2). The pixel-ring correlation inside KP12 mask 
is computed for both bands V1 \& V2, and compared with the result
from noise map (V1-V2). As we expect, the V1 and V2 results are very 
close to the V-band result in Fig. 2;
however, in the noise map result, the pixel-ring correlation coefficients 
go to nearly 0. Therefore, the pixel-ring correlation is unaffected 
by known $141\degree$ noise correlation.

\section{Latitude dependence of CMB temperatures}
The four graphs in Fig.~3 show the Galactic latitude dependence of average temperature $\<t\>$
from the foreground cleaned WMAP5 Q-, V-, W-band maps and ILC map respectively,
where for $|b|=85\degree$,
$\<t\>$ is calculated over the two regions $\pm(80\degree, 90\degree)$,
and for $|b|<80\degree$ over $\pm(|b|-2.5\degree, |b|+2.5\degree)$.
From Fig.~3 we can see systematic distortions existing in WMAP foreground cleaned
maps evidently.
The relationships of average temperature vs. Galactic latitude
in the four WMAP maps are highly consistent:   all maps have $\<t\><0$ for the nine
latitude intervals in $10\degree<|b|<55\degree$
(listed in Table~3), and, contrarily,  $\<t\>>0$
for the region of $|b|>55\degree$.
That the turning at $|b|\sim 55\degree$ and
the negativity of WMAP CMB maps in $|b|<55\degree$ can be understood
by the detected cooling effect of hot sources on their scan-rings
(shown in \S2 ans \S3)
with the fact that most of $141\degree$ rings of Galactic hot sources
are contained in the region of $|b|<55\degree$ as shown by Fig.~1.

\begin{figure}
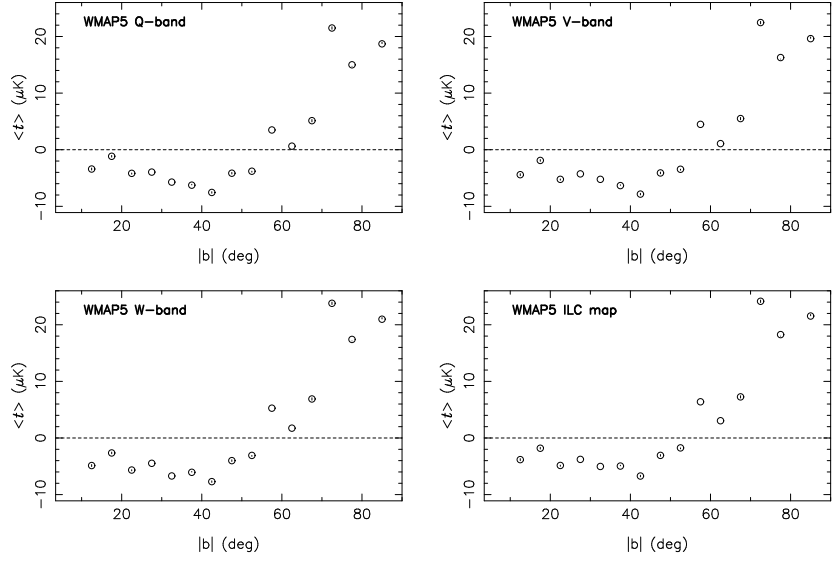

\label{f3}
    \begin{center}
     \includegraphics[width=35mm,angle=270]{f3a.ps}
 \hspace{2mm} \includegraphics[width=35mm,angle=270]{f3b.ps}\\
\vspace{3mm}
   \includegraphics[width=35mm,angle=270]{f3c.ps}
\hspace{2mm}   \includegraphics[width=35mm,angle=270]{f3d.ps}
 \caption{\footnotesize Average temperature vs.  absolute Galactic latitude
in WMAP5 maps with the Kp0 mask, from top to bottom for Q, V, W band
and ILC map respectively. The standard deviations of average temperatures are shown
in the graphs, they are small and hard to be recognized.
}
    \end{center}
\end{figure}

\begin{table}
 \caption{Average temperatures $\<t\>$ ($\mu$K) over
 Galactic latitude regions $|b|\pm2.5\degree$}
\begin{center}
\begin{tabular}{llllllllll}
\hline
  $|b|$ (deg) &  12.5 & 17.5&  22.5&  27.5&  32.5&  37.5&  42.5&  47.5&52.5\\
  \hline
  Q band & -3.4 & -1.2& -4.2& -4.0 & -5.7& -6.3& -7.6& -4.2& -3.8\\
  V band & -4.4&  -1.9& -5.2& -4.3& -5.2& -6.3& -7.8& -4.1& -3.5\\
  W band & -4.9& -2.6& -5.7 & -4.5& -6.7& -6.1& -7.7& -4.0& -3.1\\
  ILC map & -3.8  & -1.8  & -4.9 & -3.8 & -5.0 & -5.0 & -6.7 & -3.0 & -1.7 \\
  \hline
\end{tabular}
\end{center}
 \end{table}

\begin{table}
\begin{center}
\caption{Average temperatures of foreground-cleaned
WMAP5 maps over $10\degree < |b| < 55\degree$} 
\vspace{3mm}
\begin{tabular}{ l l l l l}  \hline
 Map  & Q-band & V-band & W-band & ILC \\ \cline{1-5}
$\<t\>$ ($\mu$K) & -4.6   & -5.1 & -5.3 & -4.28\\
Simulation& $0.05\pm2.1$ &$0.03\pm2.2$ &$-0.02\pm2.4$ & $0.01\pm2.0$\\ 
Significance & $ 2.2\sigma$ &$ 2.3\sigma$ & $ 2.2\sigma$  & $ 2.1\sigma$ \\
\hline
\end{tabular}
\end{center}
\vspace{-2mm}
\end{table}

\section{Discussion}
The magnitude of temperature distortion of the hottest Galactic sources upon their scan-rings can
be estimated to be $>10\mu$K from Table~1 (the corresponding region covers about $15\%$ of the
entire celestial sphere after removing KP0 mask region), and the average distortion for the region
of $|b|<55\degree$ to be $\sim 5\mu$K from Table~4 (the corresponding region covers $>80\%$ of the
entire celestial sphere or $>60\%$ after removing KP0 mask region). Such a wide-spread effect
with considerable strength can not be ignored for a precise cosmology study and need to be further
investigated.

The detected large scale distortion from CMB Gaussianity is closely connected with
the Galactic hot sources and the WMAP beam separation angle, which is hard to explain by 
the best-fit $\Lambda$CDM cosmological model  
 and most possibly comes from foreground effect and map-making.
To mitigate errors produced by effects from the hot foreground sources, the WMAP team used the Kp8
mask as a "processing mask" during the map-making process$^{[2,7]}$. However, the
detected temperature coupling between hot sources and their scan-rings indicates that the
Kp8 mask might be not wide enough to exclude all unwanted effect in the map-making process,
a broader mask, e.g. Kp0, might be better.
The future CMB observation project Planck is
designed to measure the CMB anisotropy with completely different mode to WMAP,
therefore, it is expected to be totally unaffected by such distortion.

\begin{acknowledgements}
 This study is supported by the National Natural
Science Foundation of China (Grant No. 10533020),
the National Basic Research Program of China
(Grant No. 2009CB-824800), and the Directional Research Project
of the Chinese Academy of Sciences (Grant No. KJCX2-YW-T03).
 The data analysis made use of the WMAP data archive and the
HEALPix software packages.
\end{acknowledgements}

\end{document}